\documentclass[preprint, nofootinbib]{revtex4}
\usepackage{graphicx,epsf,bm}
\usepackage{amsmath, subfigure}  
\def\be{\begin{equation}}
\def\ee{\end{equation}}
\def\beb{\begin{equation*}}
\def\eeb{\end{equation*}}
\def\bea{\begin{eqnarray}}
\def\eea{\end{eqnarray}}
\def\beab{\begin{eqnarray*}}
\def\eeab{\end{eqnarray*}}

\def\bi{\begin{itemize}}
\def\ei{\end{itemize}}

\def\w{{\omega}}
\def\cs2{c_{\rm{s}}^2}

\def \beg {\begin{enumerate}}
\def \en {\end{enumerate}}

\def\Pb{P_0}
\def\rhob{\rho_0}

\def\drho{{\delta\rho}}

\def\dPn{{\delta P_{\rm{nad}}}}

\def\cs{c_{\rm{s}}^2}

\def\p{\partial}

\def\H{{\cal H}}

\def\U0{{\bar U_0}}
\def\V1{{\bar{V_1}}}

\begin{document}
\preprint{} 
\title{The future of cosmology and the role of non-linear perturbations}
\author{Adam J.~Christopherson}
\email{a.christopherson@qmul.ac.uk}
\affiliation{Astronomy Unit, School of Mathematical Sciences,
 Queen Mary University
of London, Mile End Road, London, E1 4NS, UK.\\
}
\date{\today}

\begin{abstract}
Cosmological perturbation theory is a key tool to study the universe. The linear or first order theory
is well understood, however, developing and applying the theory beyond linear order is at the cutting 
edge of current research in theoretical cosmology. In this article, I will describe some signatures of non-linear
perturbation theory that do not exist at linear order, focusing on vorticity generation at second order. In doing so,
we discuss why this, among other features such as induced gravitational waves and non-Gaussianities, shows
that cosmological perturbation theory is crucial for testing models of the universe.
\end{abstract}

\maketitle

Over the last few decades cosmology has moved from a mainly theoretical discipline to one 
in which data is of increasing importance. This change, driven primarily by advances in technology, is
encouraging since it means that we are no longer
confined to the ``theorists' playground.'' Instead, we can test and compare specific observational signatures produced
by competing cosmological models and work towards the ultimate goal: a complete theory of the evolution of the universe.

The main observable that we have at present with which to test our theories is the Cosmic Microwave Background (CMB)
radiation.
Experiments have been performed in the years after its detection in order to obtain
more details of this radiation. For example, the Wilkinson Microwave Anisotropy
Probe ({\sc WMAP}) \cite{WMAP7}  studied the anisotropies
of the CMB, finding that it is extremely isotropic (up to one part in 100,000), and the Planck \cite{Planck}
satellite is currently taking measurements to further increase our wealth of data on the CMB. The observed
 small anisotropies are very much 
in agreement with theory, which states that quantum perturbations in the field driving inflation
can produce small primordial density fluctuations which are then
amplified through gravitational instability to form the structure that exists in the universe today.

To study the theoretical framework of the standard cosmological model 
 one uses cosmological perturbation theory. The basic idea is quite simple: we model the universe
as a homogeneous `background' Friedmann-Robertson-Walker spacetime, and build up inhomogeneous perturbations, to both the spacetime
 and its matter content, on top. The perturbations
can then be split up order-by-order, with each higher order being ``smaller'' than the one before. For example, the energy density is expanded as
\be 
\rho(x^i, \eta)=\rho_0(\eta)+\delta\rho_1(x^i, \eta)+\frac{1}{2}\delta\rho_2(x^i, \eta)+\cdots\,,
\ee
where the subscripts denote the order of the perturbation (the linear order having Gaussian statistics), and $\eta$ is conformal time.
Much work has been done to date focussing on
the linear theory, which has been successful in modelling the anisotropies in the CMB and large scale structure. Perhaps the most pioneering
work in modern cosmological perturbation theory was completed by Bardeen. In 
Ref.~\cite{Bardeen:1980kt}, Bardeen presented a systematic method for removing gauge artefacts. These
are spurious modes which occur in any relativistic perturbation theory due to the fact that the splitting into a 
background and a perturbed spacetime is not a covariant process and so there exists a non-unique mapping between
points on the background manifold and points on the perturbed manifold.
He did this by constructing gauge invariant variables, focussing on the two metric potentials $\Psi$ and $\Phi$,
which correspond to the two gauge invariant scalar metric perturbations in the longitudinal gauge.
 This work was then followed by the two review
articles by Kodama and Sasaki~\cite{ks} and Mukhanov, Feldman and Brandenberger~\cite{mfb}. These three
articles together arguably form the basis of linear metric cosmological perturbation theory.

Cosmological perturbations can be decomposed into scalar, vector and tensor perturbations and, at linear order,
 the three types of perturbations  decouple from one 
another. 
However, once we go beyond linear order, different types of perturbation no longer decouple and so, for example,
vector and tensor perturbations are sourced by scalar modes. This mathematical difference between linear
and higher orders plays an important role in the theory and can result in qualitatively 
different physics beyond linear order which will, in turn, generate different observational signatures. This is one of
 the main reasons for extending perturbation theory beyond linear order, which has been the subject of much recent
 work (see Refs.~\cite{MW2008, Malik:2008yp} 
for recent reviews and for a 
comprehensive list of references on second order cosmological perturbations).\\

As a concrete example, we will study the generation of vorticity beyond linear order. We can gain physical insight from 
classical fluid dynamics, where the vorticity, ${\bm \w}$, is defined as ${\bm \w}\equiv {\bm \nabla}\times{\bm v}$, with ${\bm v}$ the velocity
of the fluid. Using the Euler equation, one can show that the vorticity evolves as 
\be 
\label{eq:ev}
\frac{\p {\bm \w}}{\p t}={\bm \nabla}\times({\bm v}\times{\bm \w})+\frac{1}{\rho^2}{\bm \nabla}\rho\times{\bm \nabla}P\,,
\ee
for an inviscid fluid with pressure $P$ and energy density $\rho$ and in the absence of any body forces (see, e.g., Ref.~\cite{vort_special}).
 The rightmost term
is the baroclinic term and acts as a source for the vorticity. If the fluid is barotropic such that $P\equiv P(\rho)$, then this 
term vanishes. Thus, in order to obtain a source, we need to allow for a more general perfect fluid with an equation of 
state that depends not only on the energy density, but is of the form $P\equiv P(\rho,S)$, where $S$ is the entropy. This is
Crocco's theorem, which states that vorticity generation is sourced by gradients in entropy in classical fluid dynamics.

The classical fluid dynamical case hints at the potential importance of these non-linear terms in a cosmological setting.
Hence, we now investigate the cosmological
case using perturbation theory.
We define the vorticity tensor as
$
\w_{\mu\nu}={\mathcal{P}}_\mu{}^\alpha{\mathcal{P}}_\nu{}^\beta u_{[\alpha;\beta]}\,,
$
where ${\mathcal{P}}_{\mu\nu}$ projects into the fluid rest frame, and the semicolon denotes the covariant derivative. Here
$u_\mu$ is the usual fluid four velocity. By expanding
the vorticity tensor to the required order in perturbation theory we can find its evolution by taking the time derivative and using the 
governing evolution and constraint equations (from energy-momentum conservation and the Einstein equations) \cite{vorticity}. 
This gives the well known result at first order that vorticity decays in
the absence of anisotropic stress \cite{ks}. However, at second order, we obtain the following result
\be 
\label{eq:vorsecondevolution}
\w_{2ij}^\prime -3\H\cs\w_{2ij}
=\frac{2a}{\rhob+\Pb}\left\{3\H V_{1[i}\dPn_{1,j]}
+\frac{\drho_{1,[j}\dPn_{1,i]}}{\rhob+\Pb}\right\}\,,
\ee
even assuming zero first order vorticity. Here, $\delta P_{\rm nad 1}$ is the non-adiabatic pressure perturbation which 
is proportional to the entropy perturbation of the fluid\footnote{For details on the non-adiabatic pressure perturbation
in the cosmic fluid see Ref.~\cite{Brown:2011dn}.}
 ($\delta P_{\rm nad 1}\propto\delta S_1$) \cite{nonad}.
The rightmost term in this equation is the analogue of the baroclinic term in the 
classical case -- for a barotropic fluid this term vanishes, but when allowing for entropy there exists a non-zero source.
Thus we can see that this result is an extension of Crocco's theorem to an expanding, perturbed framework.
This generation of vorticity by the coupling between scalar first order perturbations is an interesting example of the new
qualitative phenomena that arise beyond linear order in perturbation theory. It will have important consequences for 
magnetic field generation \cite{Pitrou} and for B-mode polarisation of the CMB\footnote{The CMB radiation has two polarisations, the E-mode, or 
curl-free polarisation which is sourced by vector and scalar perturbations and the B-mode, or divergence-free polarisation
which is sourced by tensor and vector perturbations.} , which certainly warrants future investigation. 

Vorticity, however, is just one example of a new observational signature arising beyond linear order in perturbation theory. Another is 
the gravitational waves induced by scalar perturbations as studied, for example, in Refs.~\cite{Ananda:2006af, Baumann:2007zm}. 
As is well known, inflation is the favoured model of the early universe and generically predicts gravitational waves.
However there are other models of the early universe whose key prediction is an absence of linear tensor perturbations. 
Second order induced gravitational waves, on the other hand, are model independent, and so will exist in any model of the early universe.
Baumann et. al showed in Ref.~\cite{Baumann:2007zm} that these induced
 gravitational waves are comparable, on some scales, to the linear tensor perturbations. Therefore, one 
needs to carefully account for these before immediately ruling out a model which does not produce first order gravitational
waves. That is, the oft-acclaimed `smoking gun' of inflation may not be so definitive.

A further motivation for the increasing interest in higher order perturbation theory, independently
to the mode-coupling sources for vorticity and gravitational waves, is
 to study non-Gaussianities of the primordial perturbation. At linear order, quantities have Gaussian statistics. However when we
move beyond linear order, one gains the ability to test for non-Gaussianities induced by inflationary cosmology (for recent reviews
on inflationary non-Gaussianities see, e.g., Refs.~\cite{Wands:2010af, Langlois:2011jt}).
These signals will in principle be observable by the Planck satellite and therefore are another way in which non-linear perturbation 
theory can constrain models of the universe.\\


We still have some way to go before we achieve our goal of understanding the evolution of the universe.
 Cosmological perturbation theory has been incredibly successful in modelling the CMB and large scale structure, 
for which the linear theory is sufficient. However, in coming years we will be inundated with data and so non-linear effects, 
such as vorticity, gravitational waves and non-Gaussianity,
 will become testable. Thus, cosmological perturbation theory continues to play a crucial role 
in our quest to understand the universe in which we live.

\section*{Acknowledgements}

The author is grateful to James Lidsey and Karim Malik for fruitful discussions and useful comments.



\end{document}